\documentclass{IEEEtran}
\usepackage{paralist} \setdefaultitem{}{\textbullet }{ $\star$}{}

\usepackage[tbtags]{amsmath}
\usepackage{subfig}
\usepackage{url}
\usepackage{graphicx}

\begin{document}

\title{Your Facebook Deactivated Friend or a Cloaked Spy \\ (Extended Abstract)}
\author{ \authorblockN{ Shah Mahmood and Yvo Desmedt} \\ 
           \authorblockA{Department of Computer Science,\\
           University College London,\\
           United Kingdom\\
           Email:\{shah.mahmood, y.desmedt\}@cs.ucl.ac.uk}}

\maketitle
\begin{abstract}
With over 750 million active users, Facebook is the most famous social
networking website. One particular aspect of Facebook widely
discussed in the news and heavily researched in academic circles is
the privacy of its users. In this paper we introduce a zero day
privacy loophole in Facebook. We call this the deactivated friend
attack. The concept of the attack is very similar to cloaking in Star
Trek while its seriousness could
be estimated from the fact that once the attacker is a friend of the
victim, it is highly probable the attacker has
indefinite access to the victims private information in a cloaked way. We demonstrate the impact
of the attack by showing the ease of gaining trust of Facebook users
and being befriended online. With targeted friend requests we were
able to add over 4300 users and maintain access to their Facebook profile information for at least 261 days. No user was able to unfriend us
during this time due to cloaking and short de-cloaking sessions. The
short de-cloaking sessions were enough to get updates about the
victims. We also provide
several solutions for the loophole, which range from mitigation to a
permanent solution \footnote{This is only the extended abstract. For
  the full version of the paper please consult the proceedings of
  SESOC 2012, March 19, 2012 (IEEE
  PerCom Workshop 2012, March 19-23, 2012).}.

\end{abstract}

\begin{keywords} Facebook, Social Network, Privacy, Cloaked Channel, Cloaked:
  Nonexistent, Decloaked: Reactivated \end{keywords}

\section{Introduction}

Facebook, the popular online social network, was founded by Mark
Zuckerberg in 2004. Today it has over 750 million active users around
the globe, half of whom login to the website on daily
basis \cite{Face11}. The average user has 130 friends, spends 60 minutes on the
network every day and creates 90 pieces of content per month. 20
million new applications are installed by its users every day.  In
April 2010, Facebook launched its social plugin to integrate with
other websites. Since the launch, 2.5 million websites have integrated
into Facebook. In addition, Facebook provides a mobile platform used by
250 million people, who according to Facebook, are twice as active on
the network as the non-mobile users. In 2010,  Experian Hitwise named Facebook as
the most visited website for the second consecutive year \cite{Hitw10}. 

The large volume of data produced, and shared, the vast number of
users, limited number of employees (The ratio of Facebook users to
employees is approximately 250000:1 \cite{Face11}) and its rapid
success has not given the social
networking site enough time to take care of all its privacy violations. One
such point which is widely on the news,
heavily researched academically and a cause of concern for
Facebook officials, is the issue of Privacy of its
users. The importance of the issue could be guessed from the very fact
that a Google search, on November 7$^{th}$ 2011, returns 6.13 billion results for ``Facebook''+
``Privacy'' while it returns only 212 million results for ``Iraq''
+``war''. This simple Google search makes Facebook privacy 28.91 times more
famous (or infamous) than the Iraq war. Note that, Google search results are not
cent percent accurate \cite{McCurley10}, they give an estimated value. 

Employers have used Facebook to hire or fire employees on the basis of
their behaviour on social networks \cite{FacebookEmployee11}. Universities can screen
applicants or ensure discipline by monitoring their students on social
networks \cite{Bonneau09a}. According to a survey
by Social Media Examiner, 92\% marketers use Facebook as a tool \cite{FacebookMarketer11}. Phishers have improved their techniques by personalizing their
schemes based on the data they acquire from social networks and are
shown to be more useful than the traditional phishing schemes
\cite{Jagatic07,Polakis10}.  A woman in Indiana (US) was robbed by a
Facebook friend after she posted on her Facebook profile that she was
going out for the night  \cite{FacebookRobbery10}. 

Facebook was initially launched as a student social network,
its approach to privacy was initially network centric, which meant
all data shared by users was visible to all the members of the network
 \cite{Yardi10}. The privacy settings were
changed several times till it reached the present form, where by
default different levels of users information is visible to
``Friends'' , ``Friends of Friends'' and ``Everyone''. Today,
Facebook does allow the option of sharing the information only with
the user through ``Only Me'' option in the privacy settings, in addition to
applying exception of the rules to specific groups of people, called
``lists''. It has been shown in the literature, that users rarely
change the default settings \cite{MacKay91,Bonneau09}.

As Facebook's default settings evolved to a level where not all
information was visible to the user's network, attackers and
researchers tried to get
information from what is publicly available. Since information uploaded by the users is expected to
be shown to their friends only on Facebook \cite{Acquisti06},
information flow (and so leakage) to friends is considered authorized. We
analyze the impact of
an attacker being able to be visible on the friendlist of a user
temporarily and then hide beneath an invisibility cloak. This is the
focus of this paper. 

\section{Summary of our contributions}

\subsection{Deactivated Friend Attack} \label{OurAttack}

Our \textit{deactivated friend attack} occurs when an attacker adds their victim on
Facebook and then deactivates her own account. As deactivation is
temporary in Facebook, the attacker can reactivate her account as she
pleases and repeat the process of activating and deactivating for
unlimited number of times. While a friend is deactivated on Facebook, she
becomes invisible. She could not be unfriended (removed from friend's list) or added to any
specific list. The only privacy changes that may apply to her are
those applied to \textit{all} friends or the particular list of which it is
already a member. 

This deactivated friend i.e.\ the attacker may later reactivate the account and crawl
her victims profiles for any updated information. Once the crawling
has finished, the attacker will deactivate again. While activated, the
attacker is visible on the victim's friend list. The concept here is
very similar to that of cloaking in Star Trek where Badass Blink
or Jem'Hadar has to uncloak (be visible), even if only for a moment,
to open fire. Facebook provides no notification about the
activation or deactivation
of friends to its users. 

\subsubsection{Detection and restriction of the attacker as a friend}
As the attacker has to uncloak to spy, there is a probability
of being detected and unfriended or put under restricted privacy
policies. This probability will be dependent on several factors
including the probability the
victim checks their own friendlist ($p_1$), the probability the user
is online when the attacker is de-cloaked ($p_2$), the probability the victim
checks their profile page ($p_3$), the probability the victim checks their
friends preview (Facebook shows thumbnails and names of 10 friends on
the left side of user's profile page) 
($p_4$), the probability of the attacker being available on the
friends preview ($p_5$), the probability of the victim getting suspicious about the
attacker after finding him on the friendlist and then attempting to restrict or
unfriend her ($p_6$), the probability that the victim will be able to apply the restriction before the attacker deactivates considering
the time they both have ($p_7$) and finally
the probability of other factors ($p_8$).  For simplicity, assuming independence, the 
probability of the attacker being restricted is,
\begin{equation*} 
p_r = p_1p_2p_6p_7 + (1-p_1)\prod_{\substack{i=2}}^{8}p_i \label{eq:1}
\end{equation*}

To get a better insight into probabilities $p_1$, $p_2$, $p_3$ and $p_4$
we did a survey in which 76 people took part. In our survey,  48.5\%
participants selected 18:01-21:00 hours, while only
9.1\% selected 00:01-06:00 hours as their most likely time
to login to Facebook. The attacker may use this information and
activate only when the user is least likely to login. The time zone of
the user could be found from the current city details on the profile,
if provided, or the locations details in status messages. 79\% survey takers checked their friend list
on less than 20\% of the occasions when using Facebook. The low percentage of people
checking the friend list is justified because Facebook allows searching for the friend of interest
in the search box, so there is not much of a need to scroll down the
friend list where on average a Facebook user has 130 friends
\cite{Face11}. These friends in the list are arranged by
alphabetical order based on their first names. Any name starting with
a character close to the end of the alphabet will be less likely to
be viewed by the user. Moreover, 74.4\% users checked their profile page less than 51\%  of the
times they used Facebook. This high percentage does not
check their Facebook profile page because when a user logs in to the
social networking website, the default page is the home page. The home
page is the epicenter of most social things including the newsfeed and
events. The user is also given an option to update her status from
the home page. Moreover, any notifications give a direct link to the
comment or activity. All these things make the users own profile page
less attractive to being checked often.  Furthermore, the survey
showed 50\% people claimed to glance 20\% or less times at their
friends preview when they check their profile page. Eyetracking
analysis could give further insight to the correctness of this
percentage distribution \cite{Duchowski03}. With a
repeated use of Facebook, the user's mind is trained to search for new
information in the center of the page, where the messages, videos,
photos etc,  are posted. The scroll bar is on the righthand side, making it
a better place for the advertisements to be placed. The friends
quick view is on the lefthand side, so it is  not the most likely place
to catch attention. But attention might also depend on other factors
like the colors on the page, the details of the pictures available
etc.  A photo that could mix more with the background might not be that
attractive or eye catching. Similarly, the use of a network name under the
name of attacker might change the probability of eyes getting attracted to the
friend in the preview. Moreover, different names may have different
impacts.

Probability $p_5$ is also not very hard to estimate. Out of the 10
friends shown on the preview roughly 7 are  those with whom the user has recently interacted on
Facebook. The remaining
3 friends for the preview will be selected randomly (Facebook friend preview algorithms are
not public, these results are estimated from our observation). Having 130 friends on
average, the likelihood of the attacker ending up on the friends list
is, roughly 3/130. The Friend preview is cached by Facebook, so once the user
logs in, the same 10 friends are displayed in random order.

The attacker may use multiple pseudonyms to be present as several deactivated friends on the
victims profile so removal or identification of one of the attacker profiles might
still undermine privacy.  

\subsubsection{Seriousness of the attack} \label{AttackSerious}

The attack is very serious for several reasons. First, it is
very hard to detect this attack. The attacker can activate her account
at the moment
least likely to be detected and crawl her victims profile for
information, keeping an updated record. Various groups of information
aggregators including marketers, background checking agencies, governments,
hackers, spammers, stalkers and criminals would find this attractive as a permanent
back door to the private information of a Facebook user, once
befriended. Secondly, a user may not be privacy conscious initially
but, with the large media coverage or personal experience become
much more concerned. Then, he may want to adjust his privacy settings
on social networks like Facebook. With this attack, and the attacker
being cloaked, the victim will not be able to apply any updates, unless
they are applied to all friends, or to lists of which the attacker is a
member. Thirdly, when the attacker can closely monitor a few users on
Facebook, they can get a deeper insight into a large network. The
attacker could be a cloaked spy monitoring and analyzing
them. Facebook recently added the feature of browsing
friendships. This would help the attacker in analyzing the bond
between two of his
victims by browsing their friendship which provides information
including the month and year since when they are Facebook friends, the
events they both attended, their mutual friends, things they both
like, their photos, the messages they wrote on each others wall, 
etc. This would give a very deep insight about the level of their
relationship, the duration when they were more or less interactive, 
etc. This information could be used for several attacks including social
engineering and social phishing attacks.



\subsubsection{Being befriended at the first place}
The attacker requires to be friends with his victim in the first
place. This could be done in many different ways. The attacker
could send a random friend request and see if accepted. She could
create the profile of a famous scientist e.g., someone has made the
profile of Claude Shannon and has got 156 friends\footnote{http://www.facebook.com/claude.shannon - Checked on November
  7, 2011, Shannon's profile might be made by a fan to pay tribute but
similar approaches can be used by attackers.}. There are Facebook
profiles made for animals\footnote{The profile of a cat at the first author's
  residential hall has 159 friends, this could again be used to spy on
  users -
  http://www.facebook.com/CelebratingStarlight.}. Psychologically we do
not feel a threat from some animals, so if profiles for them are made on
Facebook then users will feel less threatened. Also, the
attacker could use social engineering by pretending to be an
acquaintance or real life friend of the victim, but not yet a Facebook
friend. Another approach would be to pretend to be someone the victim
met at a conference or during a social event.
By default, Facebook uses encryption only for login, so a man in the middle attack
or session hijacking is not very difficult. The attacker could use
this technique to take over a session, accept himself as a friend,
remove the notification of friendship (to avoid suspicion) and
then deactivate the attacking account. Other possible attacks could
include using clickjacking to be automatically accepted as a friend,
the traditional use of keyloggers, OS vulnerabilities, browser
vulnerabilities, smart phone hacks, etc. 

\subsubsection{Experimental Results}

To prove the ease of being befriended on Facebook we ran an experiment
for 606 days. The experiment was divided into three phases: the first
phase was targeting befriending of users, the second phase was getting
into the cloaked mode and temporarily de-cloaking for updates and the
third phase was permanent de-cloaking to see how many users unfriend
a stranger (whom they once added).

\paragraph{Phase 1} 

We created a Facebook account under a pseudonym, for experimental
purposes, on February 15, 2010. For the
first 99 days we requested 225 users to be friends on Facebook, out of which 90 people
accepted our request. During this time we did not allow other users to add us as
friends. We wanted to get closer to the circles of our potential
targets and did not want other random requests. On the 100th day, we
started accepting requests from
other people. With the acception of friend requests our
probability of being accepted by our desired group kept
increasing. During the course of 285 days of phase 1 of the
experiment, we sent 595 friend requests out of which 370 were
accepted, resulting in an average success rate of 62\%. We
received a total of 3969 requests from day 100 till day 285 which are
6.67 times more than the number of requests we sent. Till Day 285 we had added 4339 users as friends. This is
when we decided to move to phase 2 of the experiment.

\paragraph{Phase 2}
In phase 2 of the experiment we went into the cloaked mode with
temporary de-cloaking to get updates about our friends on
Facebook. The phase continued for a total of 261 days with de-cloaking
at regular intervals. We de-cloaked for only 10 minutes in each
session, which is enough time to crawl hundred of profiles. We were able to observe all our Facebook
friends during the de-cloaking except those who had deactivated their
accounts and were in the cloaked mode. None of our friends could technically have,
unfriended us during this phase.  

\paragraph{Phase 3}
The first and second phase of the experiment took a long time and with
the awareness about privacy increasing we expected that users will
remove strangers from their friendlist, once they are given an option. So, we left the account in
de-cloaked mode, with no activity, for 60 days. We noticed a shrink of
239 users in our friendlist. This may indicate that about 5.5\% people got
privacy conscious with the extended awareness and started removing
strangers from their friendlists.

\subsubsection{Reflections and Ethical consideration}

As mentioned in the work by Boshmaf \textit{et. al.} \cite{Boshmaf11}, very low risk experiments are
the only way to analyze the real world impact of such  vulnerabilities
in social networks used by hundreds of millions of users. 

Ethics of the research were strongly considered during the
experimental phase. We did not save any identifiable personal
information. We will apologize to all those who took part in the
study, delete the account before the end of the conference and
inform Facebook to fix the vulnerability. 

Moreover, the use of pseudonyms can be considered as a measure of
remaining anonymous which is protected by the first amendment of the
constitution of the United States, in the context of freedom of speech
e.g., 
\cite{SupremeCourt95,Kerr07}. This raises an important question whether
organizations like Facebook or, until recently, Google were following
the law when preventing users from using measures of anonymity for
the use of their respective social networks. Further discussion of the
topic is beyond the scope fo the paper.

\subsection{Solutions} \label{Solutions}

We now provide some solutions to the attack,
\\
\begin{compactitem}[\textbullet]
 \item Users should be notified about the activation and deactivation of
   their friends. Using this a user may report any suspicion to
   Facebook.    
\item Flag the users who activate and deactivate their
   accounts several times. Then they could be further monitored for
   any suspicious activities and permanently banned by
   Facebook. Alternatively, if users are notified about the activation
   and deactivation of the accounts of their friends, then a Facebook
   application can be developed which monitors such changes and
   automatically jails (reduces the view of a victims account) a
   person during the short span of de-cloaking or even unfriend them. 
 \item The deactivated friends should still be visible in the users
   friend list (may be blurred) such that the privacy settings applied to
   them maybe changed. It should also be made possible to remove them
   from the friend list. 
\item Permanent deactivation of accounts with no reactivation as is
   done in Google+, Orkut and Twitter can be another alternative. 
\end{compactitem} 

\section{Conclusion} \label{Conclusion}
In this paper we presented a privacy loophole in Facebook
in which the attacker temporarily deactivates his account to avoid
detection and removal from the friend list, but reactivates to crawl
the victims data. We also discussed the impact of the attack, the possibility
of it being detected and the possible groups of users who may be
interested in using this attack. To support our claims of ease of
targeted befriending of users on Facebook we conducted an experiment
where on average 62\% of our friend requests were accepted. We
tested the cloaking attack for 261 days and none of our
Facebook friends unfriended during the course. We
conducted a survey to get further insights into the user's behavior
and make the attack more stealthy. Finally, we provided some
recommendations about possible solutions to the problem. We wonder
whether other cloaked channels exist. 

\bibliographystyle{abbrv}
\bibliography{References_Privacy}

\end{document}